\documentclass[12pt]{iopart}

\usepackage{iopams}
\usepackage{mathrsfs}

\begin{document}

\newcommand{\A}{\mathbb{A}}
\newcommand{\R}{\mathbb{R}}
\newcommand{\C}{\mathbb{C}}
\newcommand{\Z}{\mathbb{Z}}
\newcommand{\N}{\mathbb{N}}
\newcommand{\Q}{\mathbb{Q}}
\newcommand{\OO}{\mathbb{O}}
\newcommand{\fa}{\mathfrak{a}}
\newcommand{\fb}{\mathfrak{b}}
\newcommand{\fc}{\mathfrak{c}}
\newcommand{\fd}{\mathfrak{d}}
\newcommand{\fe}{\mathfrak{e}}
\newcommand{\ff}{\mathfrak{f}}
\newcommand{\fg}{\mathfrak{g}}
\newcommand{\fh}{\mathfrak{h}}
\newcommand{\ffi}{\mathfrak{i}}
\newcommand{\fj}{\mathfrak{j}}
\newcommand{\fk}{\mathfrak{k}}
\newcommand{\fo}{\mathfrak{o}}
\newcommand{\fp}{\mathfrak{p}}
\newcommand{\fq}{\mathfrak{q}}
\newcommand{\ffl}{\mathfrak{l}}
\newcommand{\fu}{\mathfrak{u}}
\newcommand{\fv}{\mathfrak{v}}
\newcommand{\fm}{\mathfrak{m}}
\newcommand{\fn}{\mathfrak{n}}
\newcommand{\ft}{\mathfrak{t}}
\newcommand{\fs}{\mathfrak{s}}
\newcommand{\fw}{\mathfrak{w}}
\newcommand{\fx}{\mathfrak{x}}
\newcommand{\fy}{\mathfrak{y}}
\newcommand{\fz}{\mathfrak{z}}
\newcommand{\fA}{\mathfrak{A}}
\newcommand{\fB}{\mathfrak{B}}
\newcommand{\fC}{\mathfrak{C}}
\newcommand{\fD}{\mathfrak{D}}
\newcommand{\fF}{\mathfrak{F}}
\newcommand{\fG}{\mathfrak{G}}
\newcommand{\fH}{\mathfrak{H}}
\newcommand{\fI}{\mathfrak{I}}
\newcommand{\fJ}{\mathfrak{J}}
\newcommand{\fK}{\mathfrak{K}}
\newcommand{\fL}{\mathfrak{L}}
\newcommand{\fO}{\mathfrak{O}}
\newcommand{\fP}{\mathfrak{P}}
\newcommand{\fR}{\mathfrak{R}}
\newcommand{\fS}{\mathfrak{S}}
\newcommand{\fT}{\mathfrak{T}}
\newcommand{\fU}{\mathfrak{U}}
\newcommand{\fV}{\mathfrak{V}}
\newcommand{\fW}{\mathfrak{W}}
\newcommand{\fX}{\mathfrak{X}}
\newcommand{\fY}{\mathfrak{Y}}
\newcommand{\fZ}{\mathfrak{Z}}
\def\RR{\mathbb{R}}
\def\NN{\mathbb{N}}
\def\CC{\mathbb{C}}
\def\QQ{\mathbb{Q}}
\def\KK{\mathbb{K}}
\def\ZZ{\mathbb{Z}}
\def\SS{\mathbb{S}}
\def\PP{\mathbb{P}}
\newcommand{\bba}{\mathbf{a}}
\newcommand{\bb}{\mathbf{b}}
\newcommand{\bc}{\mathbf{c}}
\newcommand{\bd}{\mathbf{d}}
\newcommand{\bbe}{\mathbf{e}}
\newcommand{\bbf}{\mathbf{a}}
\newcommand{\bh}{\mathbf{h}}
\newcommand{\bk}{\mathbf{k}}
\newcommand{\bl}{\mathbf{l}}
\newcommand{\bu}{\mathbf{u}}
\newcommand{\bv}{\mathbf{v}}
\newcommand{\bw}{\mathbf{w}}
\newcommand{\bA}{\mathbf{A}}
\newcommand{\bB}{\mathbf{B}}
\newcommand{\bC}{\mathbf{C}}
\newcommand{\bD}{\mathbf{D}}
\newcommand{\bI}{\mathbf{I}}
\newcommand{\bJ}{\mathbf{J}}
\newcommand{\bL}{\mathbf{L}}
\newcommand{\bR}{\mathbf{R}}
\newcommand{\bS}{\mathbf{S}}
\newcommand{\bT}{\mathbf{T}}
\newcommand{\bU}{\mathbf{U}}
\newcommand{\bV}{\mathbf{V}}
\newcommand{\cA}{\mathcal{A}}
\newcommand{\cB}{\mathcal{B}}
\newcommand{\cC}{\mathcal{C}}
\newcommand{\cD}{\mathcal{D}}
\newcommand{\cH}{\mathcal{H}}
\newcommand{\cI}{\mathcal{I}}
\newcommand{\cJ}{\mathcal{J}}
\newcommand{\cK}{\mathcal{K}}
\newcommand{\cL}{\mathcal{L}}
\newcommand{\cM}{\mathcal{M}}
\newcommand{\cN}{\mathcal{N}}
\newcommand{\cO}{\mathcal{O}}
\newcommand{\cP}{\mathcal{P}}
\newcommand{\cQ}{\mathcal{Q}}
\newcommand{\cR}{\mathcal{R}}
\newcommand{\cS}{\mathcal{S}}
\newcommand{\cT}{\mathcal{T}}
\newcommand{\cU}{\mathcal{U}}
\newcommand{\cV}{\mathcal{V}}
\newcommand{\cW}{\mathcal{W}}
\newcommand{\cX}{\mathcal{X}}
\newcommand{\be}{\begin{equation}}
\newcommand{\ee}{\end{equation}}
\newcommand{\bea}{\begin{eqnarray}}
\newcommand{\eea}{\end{eqnarray}}
\newcommand{\nn}{\nonumber}
\newcommand{\kt}{\rangle}
\newcommand{\Br}{\langle}
\newcommand{\cun}{\mbox{\scriptsize${\cal N}$}}
\newcommand{\cum}{\mbox{\scriptsize${\cal M}$}}
\newcommand{\ed}{\end{document}}
\newcommand{\bbr}{\br\!\br}
\newcommand{\kkt}{\kt\!\kt}
\newcommand{\pbr}{\prec}
\newcommand{\pkt}{\succ}
\newcommand{\ppbr}{\prec\!\!\!\prec}
\newcommand{\ppkt}{\succ\!\!\!\succ}
\newcommand{\cbr}{(\!(}
\newcommand{\ckt}{)\!)}
\newcommand{\pleq}{\preccurlyeq}
\newcommand{\pgeq}{\succcurlyeq}
\newcommand{\rx}{{\rm x}}
\newcommand{\ry}{{\rm y}}
\newcommand{\rp}{{\rm p}}
\newcommand{\rZ}{{\rm Z}}
\newcommand{\rH}{{\rm H}}
\newcommand{\rX}{{\rm X}}
\newcommand{\rP}{{\rm P}}
\newcommand{\rh}{{\rm h}}
\newcommand{\rE}{{\rm E}}
\newcommand{\rv}{{\rm v}}
\newcommand{\rT}{M}
\newcommand{\approxN}{\stackrel{\sN}{\approx}}
\newcommand{\np}{\newpage}
\newcommand{\dl}{[\![}
\newcommand{\dr}{]\!]}
\newcommand{\sg}{{\rm sign}}
\newcommand{\xto}{\Rightarrow}
\newcommand{\dto}{\Leftrightarrow}
\newcommand{\ex}{\exists}
\newcommand{\we}{\wedge}
\newcommand{\ew}{\vee}
\newcommand{\etap}{{\eta_{_+}}}
\newcommand{\emp}{\varnothing}
\newcommand{\bce}{\begin{center}}
\newcommand{\ece}{\end{center}}
\newcommand{\sgn}{\,{\rm sgn}}
\newcommand{\RE}{\,{\rm Re}}
\newcommand{\IM}{\,{\rm Im}}

\newcommand{\One}{{\stackrel{\leftrightarrow}{1}}}
\newcommand{\Zero}{{\stackrel{\leftrightarrow}{0}}}
\newcommand{\Ep}{{\stackrel{\leftrightarrow}{\mbox{\large{$\varepsilon$}}}}}
\newcommand{\ep}{{\stackrel{\;\leftrightarrow_\prime}{\mbox{\large{$\varepsilon$}}}}}
\newcommand{\MU}{{\stackrel{\leftrightarrow}{\mbox{\large{$\mu$}}}}}
\newcommand{\Mu}{{\stackrel{\;\leftrightarrow_\prime}{\mbox{\large{$\mu$}}}}}
\newcommand{\La}{{\stackrel{\;\leftrightarrow}{\mbox{\large{$\Lambda$}}}}}
\newcommand{\Uu}{{\stackrel{\leftrightarrow}{\mbox{\large{$u$}}}}}
\newcommand{\Hh}{{\stackrel{\leftrightarrow}{\mbox{\large{$h$}}}}}
\newcommand{\Om}{{\stackrel{\leftrightarrow}{\omega}}}
\newcommand{\Ee}{{\stackrel{\leftrightarrow}{\mbox{\footnotesize{${\cal E}$}}}}}
\newcommand{\Mm}{{\stackrel{\leftrightarrow}{\mbox{\footnotesize{${\cal M}$}}}}}
\newcommand{\xA}{{\mbox{\Large{$a$}}}}
\newcommand{\xB}{{\mbox{\Large{$b$}}}}

\newcommand{\sE}{\mathscr{E}}
\newcommand{\sB}{\mathscr{B}}
\newcommand{\sF}{\mathscr{F}}
\newcommand{\sG}{\mathscr{G}}
\newcommand{\sH}{\mathscr{H}}
\newcommand{\sS}{\mathscr{S}}
\newcommand{\sV}{\mathscr{V}}

\title[Conceptual Aspects of $\cP\cT$-Symmetry and
Pseudo-Hermiticity: A status report]{Conceptual Aspects of
$\cP\cT$-Symmetry and Pseudo-Hermiticity: A status report}

\author{Ali Mostafazadeh}

\address{Department of Mathematics, Ko\c{c} University, 34450 Sar{\i}yer,
Istanbul, Turkey} \ead{amostafazadeh@ku.edu.tr}

\begin{abstract}
We survey some of the main conceptual developments in the study of
$\cP\cT$-symmetric and pseudo-Hermitian Hamiltonian operators that
have taken place during the past ten years or so. We offer a precise
mathematical description of a quantum system and its representations
that allows us to describe the idea of unitarization of a quantum
system by modifying the inner product of the Hilbert space. We
discuss the role and importance of the quantum-to-classical
correspondence principle that provides the physical interpretation
of the observables in quantum mechanics. Finally, we address the
problem of constructing an underlying classical Hamiltonian for a
unitary quantum system defined by an a priori non-Hermitian
Hamiltonian.

\end{abstract}

\maketitle

\section{Introduction}

The widespread interest in the study of non-Hermitian but ${\cal
PT}$-symmetric Hamiltonian operators such as
    \be
    H=p^2+i\,\epsilon x^3,~~~~~~~~\epsilon\in\R,
    \label{cubic}
    \ee
has its root in the observation that these Hamiltonians can actually
possess a real spectrum. This was taken as a sign that perhaps one
can relax the usual condition that the Hamiltonian operator (or more
generally observables) be Hermitian operators
\cite{bender-prl-1998}. After all, a non-Hermitian operator such as
(\ref{cubic}) would define real energy values similarly to a
Hermitian operator. This motivated the search for an ``extension of
quantum mechanics to the complex domain'' \cite{bbj-prl-2002}. There
were also claims that this extended quantum theory and its field
theoretical generalizations may provide a solution for some of the
basic problems of particle physics \cite{bbj-ajp}. Several years of
intensive research have however led to a different picture. The
purpose of the present article is to give an objective survey of
what we have really learned by studying the subject. We will base
our treatment on established facts and the basic ideas rather than
circumstantial evidence for potential usefulness of the results.
This is particularly important for the outsiders who wish to assess
the scientific merits of studying the subject and the researchers
and students who are undecided whether to join in this effort.

Among the main difficulties one encounters in studying this subject
is to make sense of imprecise mathematical statements made in some
physics literature and to deal with difficult-to-read mathematical
expositions in the relevant mathematics literature. We will try to
overcome these difficulties by ignoring the subtle issues that arise
whenever the Hilbert space is infinite-dimensional.

Throughout this article we will deal with linear operators mapping
between separable complex Hilbert spaces. A separable complex
Hilbert space $\sH$ is a complex vector space endowed with a
positive-definite inner product $\Br\cdot|\cdot\kt$ such that $\sH$
is complete as a metric space and admits a countable basis
$\{\xi_n\}$. We denote the dimension of $\sH$ by $N$, and confine
our attention to the case that $N<\infty$ unless otherwise is
obvious. Our self-imposed restriction to finite-dimensional Hilbert
spaces allows us to escape dealing with the technical issues related
to the domain of the operators. These issues can be addressed
properly using a more careful mathematical analysis that is beyond
the scope of the present article. An important observation is that
these technical problems and their resolution have no significance
as far as the basic conceptual problems of interest are concerned.

We close this section by pointing out that non-Hermitian operators
have been the subject of an extensive mathematical research that we
have no intention of reviewing in this short report. We refer the
interested readers to \cite{gohberg-krein} and references therein.
The study of physical applications of non-Hermitian operators has
also a long history. The most prominent of these is their crucial
role in the description of open quantum systems \cite{open}. There
are also numerous applications of these operators in
phenomenological/effective descriptions of a variety of physical
phenomena. Some of these involve operators with $\cP\cT$-symmetry
\cite{optics}. The present paper does not aim at presenting a
general review of non-Hermitian operators and their physical
applications. It intends to address very basic and specific
questions that arise in trying to employ these operators as
Hamiltonians for fundamental, closed, and unitary quantum systems.

\section{Hermiticity and self-adjointness}

Let $\sV$ be an $N$-dimensional complex vector space with a basis
$\{\xi_n\}$, and $H:\sV\to \sV$ be a linear operator that is
represented by the $N\times N$ matrix $\underline{H}$ in the basis
$\{\xi_n\}$. This means that the entries $H_{mn}$ of $\underline{H}$
fulfil
    \be
    H\xi_m=\sum_{n=1}^N H_{nm}\xi_n.
    \label{matrix-rep}
    \ee
$\underline H$ is said to be a \textbf{Hermitian matrix} if
    \be
    H_{mn}^*=H_{nm},
    \label{Hermitian-matrix}
    \ee
i.e., $\underline H^t=\underline H^*$ where superstript $^t$ and
$^*$ stand for transpose and complex-conjugate of $\underline H$,
respectively.

Suppose that $\underline H$ is a Hermitian matrix. Then it is
well-known that it has real eigenvalues and a complete and
orthonormal set $\{\vec e_n\}$ of
eigenvectors.\footnote{Completeness of $\{\vec e_n\}$ means that it
is a basis of $\C^N$. The orthonormality is defined in terms of the
Euclidean inner product on $\C^N$. This is given by $\Br\vec v|\vec
w\kt_E:=\vec v^*\cdot\vec w$ where the dot stands for the dot
product of vectors.} This does not however imply that the operator
$H$ possesses the same properties. In fact, we cannot even talk
about orthonormality in $\sV$ unless we endow it with an inner
product.

Now, suppose $\Br\cdot|\cdot\kt$ is an inner product on $\sV$ that
makes it into a Hilbert space $\sH$. We can use the basis vectors
$\xi_n$ to define a matrix $\underline H'$ with the entries
$\Br\xi_m|H\xi_n\kt$. The matrices $\underline H$ and $\underline
H'$ coincide provided that $\{\xi_n\}$ is an orthonormal basis of
$\sH$. If this happens to be the case and $\underline H$ is a
Hermitian matrix, then the operator $H$ does have a real spectrum
and a set $\{\psi_n\}$ of eigenvectors $\psi_n$ that forms an
orthonormal basis of $\sH$. These are the characteristic properties
of self-adjoint operators. By definition $H:\sH\to\sH$ is a
\textbf{self-adjoint operator} if for all $\phi,\psi\in\sH$ we
have\footnote{Here we ignore the domain issues. For a more precise
definition see \cite{review}.}
    \be
    \Br\phi|H\psi\kt=\Br H\phi|\psi\kt.
    \label{Hermitian-operator}
    \ee

Most textbooks on quantum mechanics follow von~Neumann's terminology
\cite{von-Neumann} of using the term ``\textbf{Hermitian operator}''
for self-adjoint operators. Whenever one has a preassigned inner
product and uses orthonormal bases to represent linear operators
there is no danger of using this termonology, because Hermitian
operators are represented by Hermitian matrices. This is the reason
why some references identify Hermitian operators with those having
Hermitian matrix representations. In the present subject, however,
it is absolutely essential not to use a basis-dependent notion such
as the Hermiticity of the matrix representation, particularly
because the basis one adopts may not be orthonormal with respect to
the physically appropriate inner product(s).

Some recent papers use the term ``Dirac Hermiticity'' of an operator
to distinguish the Hermiticity of the matrix representation (in a
non-orthonormal basis) and the Hermiticity of the operator
\cite{bender-review}. A better solution to this problem is to avoid
using matrix representation of operators as much as possible. In
particular, we define a Hermitian operator according to
(\ref{Hermitian-operator}) rather than (\ref{Hermitian-matrix}). The
former, basis-independent definition, has the advantage of
clarifying the role of the inner product in determining the
Hermiticity of a given operator. This is implicit in
(\ref{Hermitian-matrix}), because this equation implies the
Hermiticity of the operator $H$ only if the basis $\{\xi_n\}$ is
orthonormal, and this cannot be checked unless one specifies the
inner product.\footnote{One may specify the inner product by
demanding that $\{\xi_n\}$ be orthonormal. There are certain
mathematical subtleties of this procedure when $N=\infty$
\cite{review}.}

A similar situation arises in General Relativity where one has the
option of using tensor fields, that are covariant quantities, or
their component in some coordinate system, that depend on the choice
of coordinates. A simple example is the metric tensor $\mathbf{g}$
and its components $g_{\mu\nu}$ in a coordinate system that are
linked by $\mathbf{g}=g_{\mu\nu}dx^\mu\otimes dx^\nu$.\footnote{The
analogue of this equation that relates a linear operator $H$ to its
matrix representation in a basis $\{\xi_n\}$ is $H=\sum_{m,n=1}^N
H_{mn} \xi_{m}\otimes\xi^\star_n$, where $\xi^\star_n:\sH\to\cC$ is
the linear map (functional) defined by $\xi^\star_n(\sum_{m=1}^N
c_m\xi_m):=c_n$, i.e., $\{\xi^\star_n\}$ is the dual basis to
$\{\xi_n\}$, \cite{wasserman}.}

\section{A precise description of quantum systems}

Suppose that $H$ is a given linear operator acting in some complex
vector space $\sV$ and having a real spectrum and a complete set of
eigenvectors $\{\psi_n\}$. We wish to address the following central
question.
    \begin{itemize}
    \item[] \textbf{Question~1}:
Can $H$ serve as the Hamiltonian operator (or more generally an
observable) for a quantum system whose state vectors belong to
$\sV$?
    \end{itemize}
Strictly speaking this is an ill-posed question unless we make it
clear what we mean by a quantum system. Therefore, we first give a
precise definition of a quantum system.

Let $\sH_1$ and $H_1:\sH_1\to\sH_1$ be a Hilbert space and a linear
operator acting in $\sH_1$ that we call a Hamiltonian operator,
respectively. Let $(\sH_2,H_2)$ be another Hilbert space-Hamiltonian
operator pair, and $\Br\cdot|\cdot\kt_i$ denote the inner product of
$\sH_i$ for $i=1,2$. A linear operator $\cU:\sH_1\to\sH_2$ with
domain $\sH_1$ and range $\sH_2$ is said to be a \textbf{unitary
operator} if for all $\phi_1,\psi_1\in\sH_1$,
    \be
    \Br\cU\phi_1|\cU\psi_1\kt_2=\Br\phi_1|\psi_1\kt_1.
    \label{unitary=}
    \ee
We say that $(\sH_1,H_1)$ and $(\sH_2,H_2)$ are
\textbf{unitary-equivalent} if there is a unitary operator
$\cU:\sH_1\to\sH_2$ satisfying $\cU H_1=H_2\cU$, alternatively
$H_2=\cU H_1\cU^{-1}$. Unitary-equivalence is an equivalence
relation on the set of Hilbert space-Hamiltonian operator pairs. A
\textbf{quantum system} is an equivalence class\footnote{Here an
equivalence class is a set of unitary-equivalent Hilbert
space-Hamiltonian operator pairs.} $\cS$ of this equivalence
relation. Each Hilbert space-Hamiltonian operator pair belonging to
$\cS$ is called a representation of $\cS$. A quantum system $\cS$ is
said to be a \textbf{unitary quantum system} if the Hamiltonian
operator in all its representations is a Hermitian operator.

What is implicit in the above mathematical description of a quantum
system is von-Neumann's axioms of quantum mechanics. According to
these axioms, given an arbitrray representation $(\sH,H)$ of a
quantum system $\cS$, the kinematics and dynamics of a $\cS$ are
respectively determined by the Hilbert space $\sH$ and the
Hamiltonian operator $H$ in this representations. In particular, the
states and observables of $\cS$ in the representation $(\sH,H)$ are
respectively the rays (one-dimensional subspaces) of $\sH$ and
certain linear operators $O:\sH\to\sH$ acting in $\sH$. The states
are uniquely determined by the state vectors $\psi\in\sH-\{0\}$ that
in general form a dense subset of $\sH$.

The main ingredient of the kinematics of quantum mechanics is
von-Neumann's projection (measurement) axiom. Enforcing it puts a
strong restriction on the observables. Specifically, it demands that
    \begin{itemize}
    \item[]\textbf{(i)} the observables $O$ must have a complete set of eigenvectors (for
    otherwise there may be states of a quantum system that can never be
    prepared) and
    \item[]\textbf{(ii)} for every observable $O$ and state vector
    $\psi$, the expectation value, $\Br\psi|O\psi\kt/\Br\psi|\psi\kt$
    is a real number. This is because the results of measurements and
    their expected values are real numbers.
    \end{itemize}
It is this requirement of the reality of expectation values that
forces observables to be Hermitian (self-adjoint) operators acting
in $\sH$. This is a direct consequences of a well-known mathematical
theorem that is unfortunately not discussed in standard textbooks on
quantum mechanics.\footnote{See for example the appendix of
\cite{review}.} In view of this theorem the reality of the spectrum
of an operator is only a necessary condition for a consistent
implementation of the projection axiom. The Hermiticity of the
observables, however, is both necessary and sufficient. This shows
that unless one wishes to modify the projection axiom one cannot
escape the condition of the Hermiticity of observables. This also
applies to the Hamiltonian operator even at the kinematic level, if
one demands that it is also an observable of the quantum system.

The dynamics of the quantum system $\cS$ in a representation
$(\sH,H)$ is described by the Hamiltonian operator $H$ through the
Schr\"odinger (or Heisenberg) equation. Again consistency of
dynamics with projection axiom demands that the evolution of the
state vectors $\psi(t_0)\to\psi(t)=U(t,t_0)\psi(t_0)$ is affected by
a unitary operator $U(t,t_0)$. In view of the Schr\"odinger
equation: $i\hbar\frac{d}{dt}U(t,t_0)=HU(t,t_0)$, this also implies
that $H$ is a Hermitian operator.\footnote{This is known as the
Stone's theorem \cite{reed-simon}.} Therefore, a consistent
application of the projection axiom alone demands that the quantum
system must be unitary.

The von-Neumann axioms of quantum mechanics are valid in all of the
representations of a unitary quantum system. The choice of the
representation, which is clearly not unique, depends on the
observer. The freedom of making this choice is similar to an
observer's freedom to choose a particular unit system. Clearly, the
physical quantities associated with the quantum system $\cS$ are
independent of the choice of a representation. This can be shown to
be a consequence of the unitary-equivalence of the representations.
For example let $\cS$ be in a state described by
$\psi_1\in\sH_1-\{0\}$ in a representation $(\sH_1,H_1)$ and $O_1$
be an observable in this representation. The expectation value of
$O_1$ in this state is given by
$\Br\psi_1|O_1\psi_1\kt_1/\Br\psi_1|\psi_1\kt_2$. Now, let
$(\sH_2,H_2)$ be another representation of $\cS$. Then there is a
unitary operator $\cU:\cH_1\to\cH_2$ that maps $\psi_1$ and $O_1$ to
$\psi_2:=\cU\psi_1$ and $O_2:=\cU\,O_1\cU^{-1}$, respectively. In
view of (\ref{unitary=}),
$\Br\psi_2|O_2\psi_2\kt_2/\Br\psi_2|\psi_2\kt_2=
\Br\psi_1|O_1\psi_1\kt_1/\Br\psi_1|\psi_1\kt_1$. This shows that the
expectation values are representation-independent.

The above discussion of quantum systems and their representations
provides a complete answer for Question~1, namely $H$ can serve as
the Hamiltonian operator for a quantum system $\cS$ represented by
$(\sH,H)$ provided that it is a Hermitian operator acting in $\sH$.
In particular completeness of the eigenvectors and reality of the
spectrum of $H$ are necessary but not sufficient. However, it turns
out that if $H$ is not Hermitian but possesses these two properties,
then it can serve as the Hamiltonian operator for another quantum
system. As a result of a theorem established in \cite{p2}, \emph{if
$H$ has a real spectrum and a complete set of eigenvectors, one can
modify the inner product of $\sH$ to define a new Hilbert space
$\sH'$ in such a way that as a linear operator acting in $\sH'$, $H$
is a Hermitian operator.} In this way $(\sH',H)$ represents a
unitary quantum system that we denote by $\cS'$.
Refs.~\cite{p2,p3,jpa-2004} give a construction of the modified
inner product that defines $\sH'$ and consequently $\cS'$. For a
comprehensive review of this construction and related developments,
see \cite{review}.

\section{Pseudo-Hermiticity and Antilinear Symmetries}

In the preceding section we gave a complete answer to Question~1.
But this answer did not involve a discussion of $\cP\cT$-symmetry
that has in a sense become a landmark of the subject. This motivates
the following question.
    \begin{itemize}
    \item[] \textbf{Question~2}: How essential is $\cP\cT$-symmetry?
    \end{itemize}
To understand the relation between $\cP\cT$-symmetry and the idea of
modifying the inner product of the Hilbert space we need to recall
some basic mathematical notions.

A linear operator $\eta:\sH\to\sH$ is called a \textbf{pseudo-metric
operator} if it is a Hermitian automorphism, i.e., it is a Hermitian
one-to-one linear operator having $\cH$ both as its domain and
range. A \textbf{metric operator} is a positive-definite
pseudo-metric operator.

We say that a linear operator $H:\sH\to\sH$ is
\textbf{pseudo-Hermitian } if there is a pseudo-metric operator
$\eta:\sH\to\sH$ satisfying\footnote{Here $H^\dagger$ denotes the
adjoint of $H$ that is defined by the condition:
$\Br\phi|H\psi\kt=\Br H^\dagger\phi|\psi\kt$ for all
$\phi,\psi\in\sH$. For a more general and precise definition of the
adjoint operator, see \cite{review}.}
    \be
    H^\dagger=\eta\,H\,\eta^{-1}.
    \label{ph}
    \ee
Suppose we are given a particular pseudo-metric operator
$\eta:\sH\to\sH$. Then a linear operator $H$ satisfying (\ref{ph})
is called \textbf{$\eta$-pseudo-Hermitian}.

Given a pseudo-Hermitian operator $H$, one can choose one of the
pseudo-metric operators $\eta$ satisfying (\ref{ph}) to construct a
pseudo-inner product according to
    \be
    \Br\phi|\psi\kt_{_\eta}:=\Br\phi|\eta\psi\kt,
    \label{pseudo-inn}
    \ee
where $\Br\cdot|\cdot\kt$ is the inner product of $\sH$. If $\eta$
happens to be a positive-definite operator, then
$\Br\cdot|\cdot\kt_{_\eta}$ is a genuine positive-definite inner
product and we can use it to define a Hilbert space $\sH'$ in which
$H$ acts as a Hermitian operator. If a positive-definite $\eta$
fulfilling (\ref{ph}) exists, $H$ is called \textbf{quasi-Hermitian}
\cite{quasi}.

It turns out that a necessary and sufficient condition for
pseudo-Hermiticity of a linear operator with a complete set of
eigenvectors is that it commutes with an invertible antilinear
operator \cite{p3,jmp-2003}. In particular, $H$ is quasi-Hermitian
if and only if it has a complete set of common eigenvectors with an
invertible antilinear operator. This is the link to
$\cP\cT$-symmetry. Because $\cP\cT$  is just a particular example of
an invertible antilinear operator, $\cP\cT$-symmetric
quasi-Hermitian Hamiltonians operators constitute a special class of
quasi-Hermitian operators. This shows that indeed $\cP\cT$-symmetry
is not an essential ingredient of the subject. One can easily
construct quasi-Hermitian Hamiltonian operators that possess other
types of invertible antilinear symmetries \cite{jpa-2008a}. One can
apply the procedure of defining a unitary quantum system by
modifying the inner product of the Hilbert space using these
Hamiltonian operators. Typical examples are the complex point
interactions \cite{jpa-2006b,jpa-2009}.

\section{Correspondence principle and classical limit}

In section~3, we outlined a mathematical description of a quantum
system $\cS$ in terms of unitary-equivalence classes of Hilbert
space-Hamiltonian operator pairs $(\sH,H)$. An important aspect of
this formulation is the procedure according to which we assign a
physical meaning to observables. This is essentially based on the
quantum-to-classical correspondence principle.

Consider a typical quantum system represented by $(\sH_0,H)$ where
$\sH_0$ is the space of square-integrable functions,
$L^2(\R):=\{\psi:\R\to\C~|~\int_\R|\psi(x)|^2dx<\infty\}$, endowed
with the standard $L^2$-inner product,
$\Br\phi|\psi\kt:=\int_\R\phi(x)^*\psi(x)dx$. It is customary to
take the operators $\hat x,\hat p:L^2(\R)\to L^2(\R)$ defined by
    \be
    (\hat x\psi)(x)=x\psi(x),~~~~
    (\hat p\psi)(x)=-i\hbar\frac{d}{dx}\psi(x),
    \label{x-hat}
    \ee
as the position and momentum observables in this representation. The
assignment of the physical meaning of ``position'' and ``momentum''
to purely mathematical entities such as $\hat x$ and $\hat p$ is a
manifestation of the quantum-to-classical correspondence
principle/or the canonical quantization scheme. For the case we
consider, this principle assigns the operators $\hat x$ and $\hat p$
to the classical observables of position and momentum of a particle
moving on the real line. It is absolutely essential that this
assignment is consistent with the correspondence of the Poisson
brackets with commutators: $ \{\cdot,\cdot\}\leftrightarrow
    (i\hbar)^{-1}[\hat\cdot,\hat\cdot]$.

Now, consider modifying the $L^2$-inner product as follows. Let
$\eta:=e^{-\kappa \cP}$ where $\kappa\in\R$ and $\cP$ is the parity
operator: $(\cP\psi)(x):=\psi(-x)$. $\eta$ is a metric operator and
$\Br\cdot|\cdot\kt_{_\eta}$ is a positive-definite inner product on
$L^2(\R)$. It is easy to check that the free particle Hamiltonian
$H_0:=\hat p^2/(2m)$ is $\eta$-pseudo-Hermitian. Therefore, if we
define $\sH'$ by endowing $L^2(\R)$ with the inner product
$\Br\cdot|\cdot\kt_{_\eta}$, we find a quantum system represented by
$(\sH',H_0)$. As operators acting in $\sH'$, $\hat x$ and $\hat p$
are not Hermitian. In particular, there is no justification for
calling them position and momentum of a particle. What plays the
role of $\hat x$ and $\hat p$ in $\sH'$ are the operators: $\hat
x':=e^{\kappa\cP}\hat x$ and $\hat p':=e^{\kappa\cP}\hat p$,
\cite{jpa-2006a}. In other words the quantum-to-classical
correspondence principle takes the form:
    \be
    x\leftrightarrow \hat x',~~~~
    x\leftrightarrow \hat x',~~~~
    \{\cdot,\cdot\}\leftrightarrow
    (i\hbar)^{-1}[\hat\cdot,\hat\cdot].
    \label{cp}
    \ee
It is not difficult to show that in fact $(\sH_0,H_0)$ and
$(\sH',H_0)$ are unitary-equivalent. Therefore $(\sH',H_0)$ is just
another equally admissible representation of the quantum system
consisting of a free particle moving on $\R$.

Now, consider a more general case where $(\sH_0,H)$ does not
represent a unitary quantum system, but modifying the inner product
of $L^2(\R)$ we obtain a Hilbert space $\sH'$ such that $(\sH',H)$
represents a unitary quantum system $\cS$. A typical example is the
$\cP\cT$-symmetric Hamiltonian~(\ref{cubic}). In this case we cannot
assign any physical meaning to Hermitian operators acting in
$\sH_0$. Rather, we need to construct Hermitian operators acting in
$\sH'$ and set up a correspondence between these and the classical
observables. For the explicit form of the position and momentum
operators associated with the unitary quantum systems that are
determined by the Hamiltonian (\ref{cubic}), see \cite{jpa-2006a}.

Once the operators $\hat x'$ and $\hat p'$ associated with position
and momentum observables in the representation $(\sH',H)$ are
determined, we can express the Hamiltonian operator $H$ in terms of
$\hat x'$ and $\hat p'$ and take the classical limit:
    \be
    \hat x'\to x,~~~~~\hat p'\to p,~~~~~\hbar\to 0.
    \label{class}
    \ee
This yields the underlying classical Hamiltonian for the unitary
quantum system $\cS$. It is via this procedure that we can give a
physical meaning to $H$ and consequently $\cS$.

A more convenient method of determining the underlying classical
Hamiltonian is by constructing a Hermitian Hamiltonian operator $h$
acting in $\sH_0$ such that $(\sH_0,h)$ is unitary-equivalent to
$(\sH',H)$. We can obtain $h$ using the following formula provided
that we are given a metric operator $\eta$ such that $H$ is
$\eta$-pseudo-Hermitian.
    \be
    h:=\eta^{1/2}H\eta^{-1/2}.
    \label{h}
    \ee
$h$ is a Hermitian operator acting in $\sH_0$, because
$\eta^{1/2}:\sH'\to\sH_0$ is a unitary operator, \cite{jpa-2003}.
Having obtained the representation $(\sH_0,h)$ we have the standard
choice for position and momentum operators and can identify the
underlying classical Hamiltonian by expression $h$ in terms of $\hat
x$ and $\hat p$  and taking the usual classical limit:
    \be
    \hat x\to x,~~~~~\hat p\to p,~~~~~\hbar\to 0.
    \label{class-usu}
    \ee
The existence of the representation $(\sH_0,h)$  seems to indicate
that we can completely avoid the use of nonstandard inner products
and apply the standard methods of quantum mechanics to describe the
quantum system $\cS$. This is true in principle, but extremely
difficult to implement in practice. The reason is that unlike $H$,
the equivalent Hermitian Hamiltonian $h$ is, in general, a highly
nonlocal (integral) operator.

In Ref.~\cite{bender-jmp-99}, the authors make another proposal for
assigning an underlying classical system for the unitary quantum
systems defined by the quasi-Hermitian Hamiltonians such as
(\ref{cubic}). This involves implementing the usual classical limit
(\ref{class-usu}) in the expression for $H$ directly and imposing
the Hamilton's classical equations of motion that correspond to the
resulting classical Hamiltonian $\fH$. The main difficulty with this
approach is that $\fH$ is a complex-valued function of $x$ and $p$.
As a results, the Hamilton's equations define a classical dynamical
system in the complex phase space $\C^2$ rather than the real phase
space $\R^2$ (that have $(x,p)$ as its coordinates.)

A careful study of the structure of this complex dynamical system
reveals that its dynamics is not consistent with the usual Poisson
bracket (symplectic structure) on the phase space $\C^2=\R^4$. To
assure the dynamical-kinematical consistency of the description of
this system, one is forced to endow the phase space $\C^2=\R^4$ with
a modified Poisson bracket \cite{CM-jmp-2007,pla-2006}. It turns
that these complex classical systems also admit a real description
and using this real description one discovers that they are
completely integrables systems possessing a specific gauge symmetry
\cite{pla-2006,smilga}.

A major problem with identifying these complex classical dynamical
systems with the classical counterparts of the original unitary
quantum systems is the lack of an explicit quantum-to-classical
correspondence (such as (\ref{cp})). In fact, such a correspondence
cannot be implemented directly, because the quantum system has a
two- (real) dimensional phase space whereas the complex classical
system has a four- (real) dimensional phase space. One may attempt
to reduce the phase space to a two-dimensional subspace by fixing a
gauge. This has so far not led to a desired correspondence between
quantum and classical observables. More problematic is that even for
non-unitary quantum systems one can apply the same method and obtain
a complex classical system.

Recently Bender et al \cite{bender-et-al} have tried to restrict
this complex classical dynamics to certain contours in the complex
$x$-plane and introduce a real, positive, and integrable function on
these contours that they propose to identify with a probability
density. In the opinion of the present author, one cannot begin to
speak about a probability density before making it clear which
observable one is measuring and what kind of a measurement axiom one
adopts. All this cannot be establish before one devises a
correspondence rule between classical and quantum observables.

\section{Concluding Remarks}

The advent of non-Hermitian $\cP\cT$-symmetric Hamiltonians with a
real spectrum led to the expectations that one can indeed extend
quantum mechanics to a more general physical theory in which such
Hamiltonian operators can also be used to model fundamental unitary
quantum systems. Our current understanding is that we achieve the
latter goal not by modifying or extending quantum mechanics as a
physical theory but by using alternative representations of quantum
systems where the Hilbert space is defined by a nonstandard inner
product. This is realized within the confines of the standard
quantum mechanics provided that we give a precise and sufficiently
general definition of a quantum system.

It turns out that actually $\cP\cT$-symmetry does not play an
essential role in implementing this idea. It serves as a particular
manifestation of the mathematical fact that every linear operator
that is capable of serving as the Hamiltonian operator $H$ for a
unitary quantum system commutes with an invertible antilinear
operator $\fS$. In fact, $H$ and $\fS$ share a common complete set
of eigenvectors.

The unitary quantum systems defined through the modification of the
inner product of the Hilbert space cannot be given a physical
interpretation unless one specifies their underlying classical
Hamiltonian. We outlined the existing methods of achieving this and
commented on the necessity of modifying the symplectic structure on
the phase space of the complex dynamical systems obtained by taking
the standard classical limit of the non-Hermitian Hamiltonians.

We conclude by emphasizing that the main problems related with the
conceptual and structural aspects of the subject have more or less
been resolved. What remains to be investigated are the concrete
physical applications of the results. Among interesting developments
in this direction are the applications in relativistic quantum
mechanics \cite{kg,prola}, quantum cosmology \cite{qc}, quantum
field theory \cite{qft}, bound-state scattering \cite{matzkin}, and
electromagnetic wave propagation \cite{epl-2008}.

\end{document}